%arXiv version

\magnification=1200.

%6-21-22

\def\qm{quantum mechanics}
\def\Qm{Quantum mechanics}
\baselineskip= 14.5pt
\parskip = 5pt
\parindent = 18pt

\centerline{\bf A NOTE ON THE QUANTUM MEASUREMENT PROBLEM}
\bigskip

\centerline{N. David Mermin}

\medskip
{\narrower\narrower

{\noindent A slightly revised version of what appeared as a ``Quick Study'' in {\it Physics Today\/}, June 2022, 63-64.}

}

\medskip 
\centerline{******}
\centerline{\bf There Is  No Quantum Measurement Problem}

 {\narrower\narrower\narrower
 
\noindent 
 {\sl The idea that wave-function collapse is a physical process stems from a %widespread 
misunderstanding of  probability and the role it plays in  quantum mechanics.}

}
 
\bigskip

There are three types of quantum physicists:  (1) those who think quantum mechanics is defaced by a so-called measurement problem;  (2) those who think, as I do, that there is no measurement problem; and (3) those who think  the issue is not worth serious thought.  You can find the diverse  views of 17 
physicists and philosophers from the first two groups in Chapter 
7 of Maximilian Schlosshauer's {\it Elegance and Enigma.\/}

Most people in all three groups would agree on the following:
Quantum mechanics  describes a physical system entirely in terms of  states.  A  state is a compendium of probabilities of all  possible answers to all  possible questions one can ask of the system.   \Qm\ is inherently statistical.  There is no deeper underlying theory that gives a fuller description.  
 
 The state assigned to a system can change in time in two ways.   
 If  no  question is asked of a system, then its state evolves in time deterministically: continuously and according to fixed rules.     
 If a  question is asked of a system ---  called making a  measurement --- then when the question is answered the state changes discontinuously into a state that depends both on the  state just  before the question was asked  and on the particular answer the system gives to that question.   This second process is  called the collapse of the state.  
Collapse is generally abrupt, discontinuous, and stochastic.

A physical system together with  another physical system that carries out a particular measurement --- an apparatus --- can be treated  by quantum mechanics as a single composite system.  If the composite system is not questioned then quantum mechanics gives a deterministic time evolution to the state assigned to it.  If the entire composite  system  is questioned, however,  the state assigned to the composite system gives probabilities  that correlate the possible answers given by the state assigned to the original system with states assigned to the apparatus that indicate those  possible answers.   The associated probabilities are just those that quantum mechanics would give for the original system alone.      So as far as probabilities are concerned, it makes no difference whether one applies quantum mechanics  to the original system alone, or to the composite original-system + apparatus. 

Many physicists in group 2 would add the following:
There are no consequences of a quantum state assignment other than all the probabilities it gives rise to.  While many (perhaps most) physicists view probabilities as objective features of the world, 
most probabilists or statisticians do not.   As the celebrated probabilist Bruno de Finetti put it in 1931,  ``The abandonment of superstitious beliefs about the existence of Phlogiston, the cosmic ether, absolute space and  time$\ldots\,$, or Fairies and Witches, was an essential step along the road to scientific thinking.  Probability too, if regarded as something endowed with some kind of objective existence, is no less a misleading misconception, an illusory attempt to exteriorize or materialize our actual probabilistic beliefs.''

Physicists  who do materialize their own probabilistic beliefs must also materialize  quantum states, which are nothing more than catalogs of such beliefs.  But a physicist who regards probabilities as personal judgments must necessarily view the quantum states he or she assigns as catalogs of his or her own personal judgments.   This view that the quantum state of a system expresses only the belief of the particular physicist who assigns that state to that system, was emphasized  as the crucial key to the interpretation of quantum mechanics by Carlton Caves, Christopher Fuchs, and R\"udiger Schack at the turn of the 21st century
\vskip 5pt
\noindent{\bf The quantum measurement problem}

The measurement problem
stems from the two different ways of viewing a measurement: the system alone, or the system + apparatus.      If the system alone is measured its state collapses.   But the state of the composite system + apparatus does not collapse until the apparatus is examined.   Which description is  correct?  Which is the real state?  

The answer from group (2) is that there is no real state of a physical system.   What one chooses to regard as the physical system, and what state one chooses to assign to it, depend on the judgment of the particular physicist who questions the system and who uses  \qm\  to calculate the probabilities of the answers.
 
This interplay between continuous and stochastic time evolution is also a feature of ordinary classical probability.   When a statistician assigns a probability to the answers to questions about a system, those probabilities vary in time by rules giving the smooth time evolution of the the isolated unquestioned system.   But those probabilities also depend on any further information the statistician acquires about the system from any other source.  That updating  of probabilities is the abrupt and discontinuous part of the classical process.   Nobody has ever worried about a classical measurement problem.  

 If the entire content of a quantum state is the catalog of probabilities it gives rise to, then
each physicist using quantum mechanics is acting as a statistician.   The acquisition of further information by that physicist --- whether it be through reading the display of an apparatus, or through communication with other physicists, or just through rethinking what that physicist already knows --- can lead to an abrupt change in those probabilities, and thus to an updating of the quantum state which the physicist uses to represent them.   There is 
no quantum measurement problem.    

Physicists in group 1 deal with their measurement problem in a variety of ways: 

In their otherwise superb quantum mechanics text
Landau and Lifshitz insist that quantum mechanics is not to be viewed as a conceptual tool used by observers.  This leads them to declare that a measurement is an interaction between objects of quantum and classical  types. How to distinguish between these two types  (which they never explain) is their (unstated) measurement problem.

Others eliminate the physicist from the story by introducing a particular kind of physical noise that interacts significantly only with subsystems that contain macroscopically many degrees of freedom. The special noise is designed to provide a physical mechanism for an objective collapse of an objective state.  They solve their measurement problem by introducing a new physical processes.

Still others remove the personal  judgment of each physicist by eliminating collapse entirely. They take quantum states to describe an inconceivably vast multitude of continuously bifurcating universes (the many-worlds interpretation) that contain every possible outcome of every possible measurement.

Such solutions all take quantum states to be objective properties of the physical system they describe and not as catalogs of personal judgments about those physical systems made by each individual user of quantum mechanics. 

\vskip 5pt

\noindent{\bf Keep the scientist in the science}

Why does our understanding of scientific laws have to be impersonal? Science is a human activity. Its laws are formulated in human language. As empiricists most scientists
believe that their understanding of the world is based on their own personal experience.
Why should I insist that {\it my\/}  interpretation of science, which {\it I \/} use to make sense of 
the world that {\it I\/} experience,  
should never make any mention of  {\it me\/}?
The existence of a ``quantum measurement problem'', either unsolved or with
many incompatible solutions, is powerful evidence that the experience of the scientist
does indeed play as important a role in understanding quantum theory, as the experience of the statistician plays in understanding ordinary probability theory.

Many physicists dismiss this view with the remark that quantum states were collapsing in the early universe, long before there were any physicists.   I wonder if they also believe that probabilities were updating in the early universe, long before there were any statisticians.

Niels Bohr never mentions a quantum measurement problem.   I conclude 
with a  statement of his that concisely expresses the above view that there is no such problem, {\it provided\/} both occurrences of  ``our'' are read not as all of us collectively but as each of us individually.  
``In our description of nature the purpose is not to disclose the real essence of the phenomena but only to track down, so far as it is possible, relations between the manifold aspects of our experience."   I believe that this unacknowledged ambiguity of the first person plural lies behind much of the misunderstanding that still afflicts the interpretation of quantum mechanics.

%Of course isolated excerpts from Bohr can support many diverse views.   But a quarter century after publishing the above he wrote that ``physics is to be regarded not so much as the study of something {\it a priori\/} given, but as the development of methods for ordering and surveying human experience.''    It's the same opinion, and it has the same ambiguity: Is ``human experience'' individual or collective?  
\bigskip
\noindent{\bf Additional resources}
\vskip 2pt

\noindent    M.~Schlosshauer, ed., {\it Elegance and Enigma: The Quantum Interviews\/},  Springer, 2011, Chap. 7.

\noindent        B.~De Finetti, {\it Theory of Probability\/} Interscience, 1990, Preface.  (Translation of {\it Probabilismo\/}, Logos {\bf 14} (Napoli) 163-219 (1931).)

\noindent     C.~A.~Fuchs  and R. Schack,  {\it Quantum-Bayesian Coherence\/},
Reviews of Modern Physics {\bf 85}, 1693 (2013).

\noindent   N.~D.~Mermin,  {\it Making Better Sense of Quantum Mechanics,\/} Reports on Progress in Physics, 
{\bf 82}, 012002 (2019).

\noindent   N. Bohr, {\it Atomic Theory and the Description of Nature\/}, Cambridge U. Press, 1934, p. 18.
%{\it Essays 1958-1962 on Atomic Physics and Human Knowledge\/}, Ox Bow Press, p 10.

\bye